\documentclass[pre,twocolumn,showpacs]{revtex4}
\usepackage{epsfig}

\begin{document}

\title{``Breathing'' rogue wave observed in numerical experiment} 
\author{V.~P. Ruban}
\email{ruban@itp.ac.ru}
\affiliation{Landau Institute for Theoretical Physics,
2 Kosygin Street, 119334 Moscow, Russia}

\date{\today}

\begin{abstract}
Numerical simulations of the recently derived fully nonlinear equations 
of motion for weakly three-dimensional water waves
[V.P. Ruban, Phys. Rev. E {\bf 71}, 055303(R) (2005)] 
with quasi-random initial conditions are reported, which
show the spontaneous formation of a single extreme wave on the deep water.
This rogue wave behaves in an oscillating manner and 
exists for a relatively long time (many wave periods)
without significant change of its maximal amplitude.
\end{abstract}


%47.15.km Potential flows
%47.35.Bb Gravity waves
%47.11.-j Computational methods in fluid dynamics

%47.10.-g General theory in fluid dynamics
%47.10.Df Hamiltonian formulations
%47.11.Kb Spectral methods
%47.15.K- Inviscid laminar flows
%47.35.-i Hydrodynamic waves 
%47.27.De Coherent structures

\maketitle

%%%%%%%%%%%%%%%%%%%%%%%%%%%%%%%%%%%%%%%%%%%%%%%%%%%%%%%%%%%%%%%%%%%%%%%%%%%

\section{Introduction}

The rogue waves (rare extreme events on a relatively calm sea surface,
alternatively called freak, killer, or giant waves),
for a long time a part of marine folklore, since 1970's have been methodically
documented by oceanographers (see review \cite{Kharif-Pelinovsky} 
for examples and some relevant statistics). 
From the viewpoint of  nonlinear science, a rogue wave is an
extremely nonlinear object --- typically, 
the amplitude of a freak wave in maximum
is close to the amplitude of the corresponding limiting Stokes wave, that is
$h/\lambda\approx 0.10\dots0.14$, where $h$ is the peak-to-trough height, 
and $\lambda$ is the length of the wave \cite{DZ2005Pisma}.
Therefore, for adequate quantitative investigation, this phenomenon requires
fully nonlinear equations and accurate numerical methods. 
For two-dimensional (2D) potential flows with a free boundary, a very efficient 
numerical scheme has been developed recently by Zakharov and co-workers 
\cite{ZDV2002}. The scheme is 
based on exact (1+1)-dimensional equations of motion written for the surface 
shape and for the boundary value of the velocity potential
in terms of the so called conformal variables
(the corresponding exact 2D theory is described in
Refs.~\cite{DKSZ96,DZK96,DLZ95,D2001,CC99,R2004PRE,R2005PLA}). 
The method extensively uses algorithms of the discrete fast Fourier transform (FFT). 
With applying this method, impressive computations have been
performed, where a numerical giant wave developed due to the Benjamin-Feir 
(modulational) instability \cite{Benjamin-Feir,Zakharov67} 
from a slightly perturbed Stokes wave. 
The spatial resolution in these numerical experiments was up to 
$2\cdot10^6$ points \cite{DZ2005Pisma}.
As to three-dimensional (3D) flows, unfortunately, a similar exact and compact
(2+1)-dimensional form of equations  is absent. Therefore ``exact''
3D simulations are currently based on the rather expensive boundary element
method (BEM) and its modifications (see 
\cite{Clamond-Grue-2001,Fructus_et_al_2005,Grilli,GuyenneGrilli2006}, 
and references therein). Since the underlying algorithms of BEM 
are quite complicated, the best practically achieved spatial resolutions 
on the free surface for essentially 3D waves are typically few tens multiplied
by few tens, as in the recent works 
\cite{Fructus_et_al_2005,Grilli,GuyenneGrilli2006}. Definitely, this is not 
sufficient to simulate large wave systems with dozens and hundreds waves,
as it is necessary for practical applications. 
We exclude here the approximate equations 
describing wave envelopes \cite{Dysthe1979,TKDV2000,OOS2000,Janssen2003},
because they are not appropriate in real situations when many random
waves with very different wave vectors and amplitudes are excited. 
Other approximate equations, for instance the weakly nonlinear Zakharov 
equations \cite{Z1999,OOSRPZB2002,DKZ2004,LZ2005}, 
are only good if the wave steepness is small, 
but this is clearly not the case for the extreme waves. However,
though rogue waves are strongly nonlinear, and the wave steepness cannot serve
anymore as a small parameter of the theory, 
nevertheless another small parameter may exist in the system. 
Namely, practically important is the situation when relatively long 
(much longer than a typical wave length) wave crests are 
oriented along a definite horizontal direction. 
For such weakly 3D flows, the advantages of the conformal variables 
are almost recovered, as it has been explained in 
Refs.~\cite{R2005PRE, RD2005PRE}. In the cited papers, 
the noncanonical Hamiltonian description in terms of the conformal variables has
been generalized from 2D to 3D potential inviscid flows with a free surface, and
the asymptotic expansion of the Hamiltonian functional on the small parameter 
$\epsilon=(l_x/l_q)^2$ has been suggested, where $l_x$ is a typical 
wave length, and $l_q$ is a large transversal scale along the wave crests.
In particular, the first-order 3D corrections have been calculated explicitly. 
What is important, all linear operators coming into the equations are diagonal 
in the Fourier representation. Therefore a relatively high spatial 
resolution ($16384\times 256$ in Ref.\cite{RD2005PRE})
for the corresponding numerical algorithm has been possible due
to the large number of local operations that result from the Fourier
diagonalization. In Ref.~\cite{RD2005PRE} some numerical results 
have been presented, for non-random initial conditions and 
typical dimensionless wave numbers about 20.

\begin{figure}
\begin{center}
   \epsfig{file=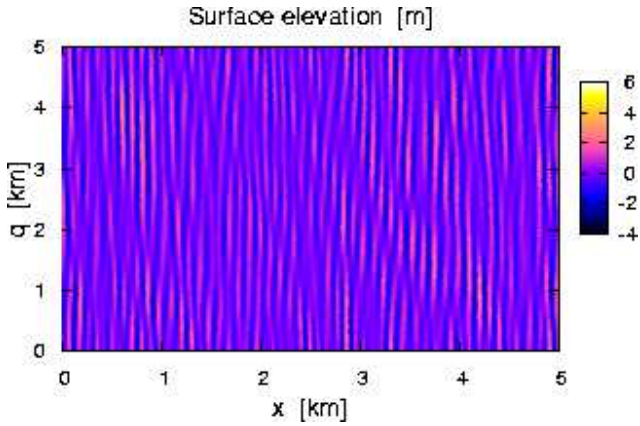,width=84mm}  
\end{center}
\caption{(Color online). 
Map of the free surface at $t=0$.} 
\label{t0-map}
\end{figure}

\begin{figure}
\begin{center} 
  \epsfig{file=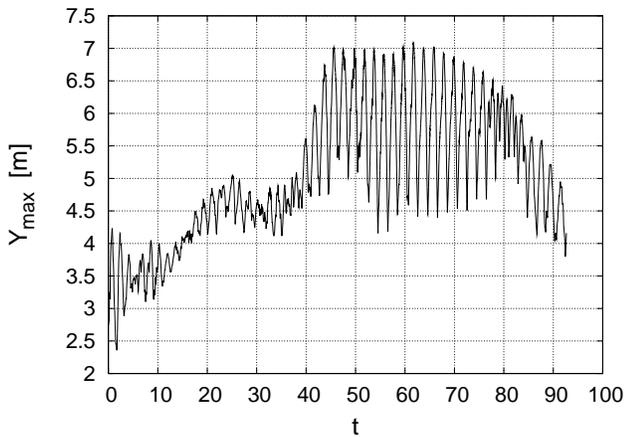,width=85mm}       
\end{center}
\caption{Maximum wave height versus dimensionless time.} 
\label{Ymax}
\end{figure}

\begin{figure}
\begin{center} 
  \epsfig{file=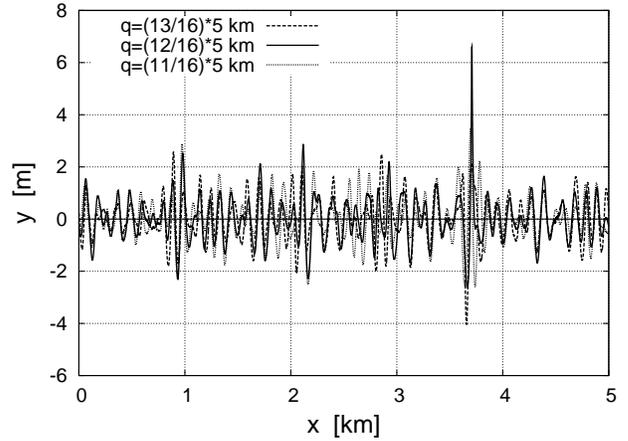,width=85mm}        
\end{center}
\caption{Wave profiles at $t=50$.} 
\label{t2-profiles-long}
\end{figure}

In the present work another numerical experiment is reported, which is
more close to reality. Main wave numbers now are about 50, and the computations 
start with a quasi-random initial state (shown in Fig.~\ref{t0-map}). 
Concerning efficiency of the numerical implementation, it should be noted that 
with the FFTW library \cite{FFTW3}, it takes
less than 2 min to perform one step of the Runge-Kutta-4 numerical integration 
on an Intel Pentium 4 CPU 3.60GHz with 2048M memory, 
for the maximal possible spatial resolution $16384\times 512$.
Here a giant wave formation has been observed as well, but contrary to the
previous computations \cite{DZ2005Pisma} and \cite{RD2005PRE}, 
this freak wave is not breaking, but it exists for many wave periods without
tendency towards increasing or decreasing its maximal amplitude (which 
in this case is distinctly less than the limiting Stokes wave amplitude,
see Figs.~\ref{Ymax}-\ref{t2-profiles-long}). 
During the life time, the rogue wave behaves in an oscillating manner, with the
highest crest being alternately ahead or behind of the deepest trough.
Observation of such kind of behavior is important for better
understanding of the rogue wave phenomenon.

\section{Equations of motion}
%{\it Equations of motion.---}

\begin{figure}
\begin{center}
  \epsfig{file=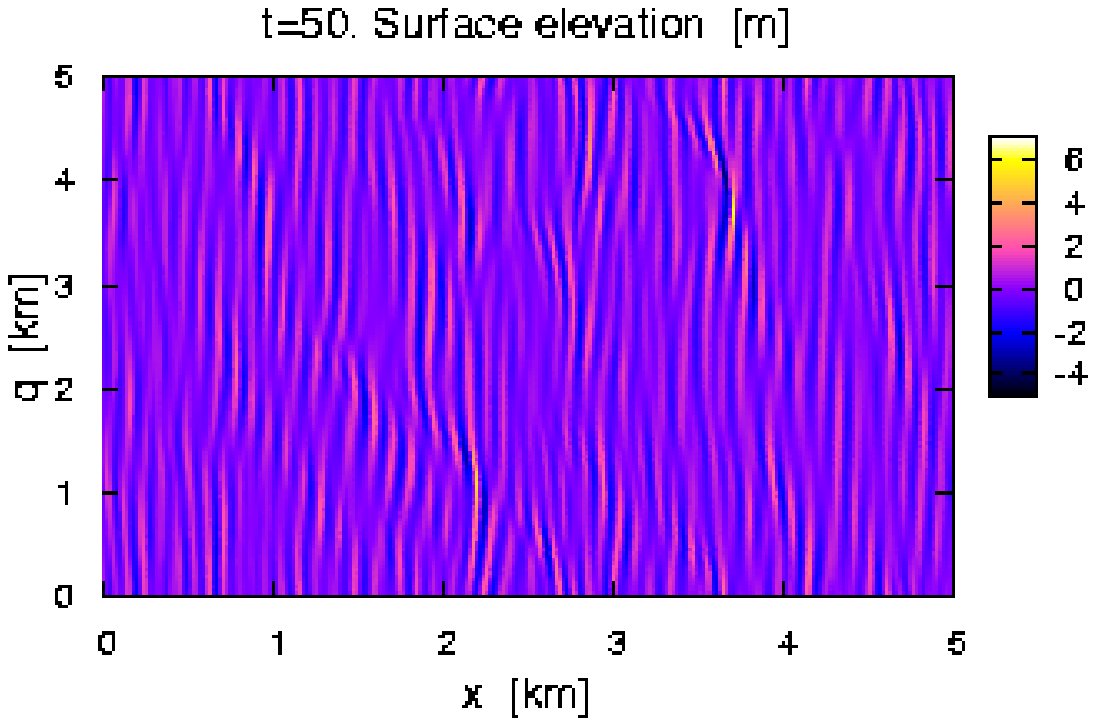,width=84mm}   
\end{center}
\begin{center}
  \epsfig{file=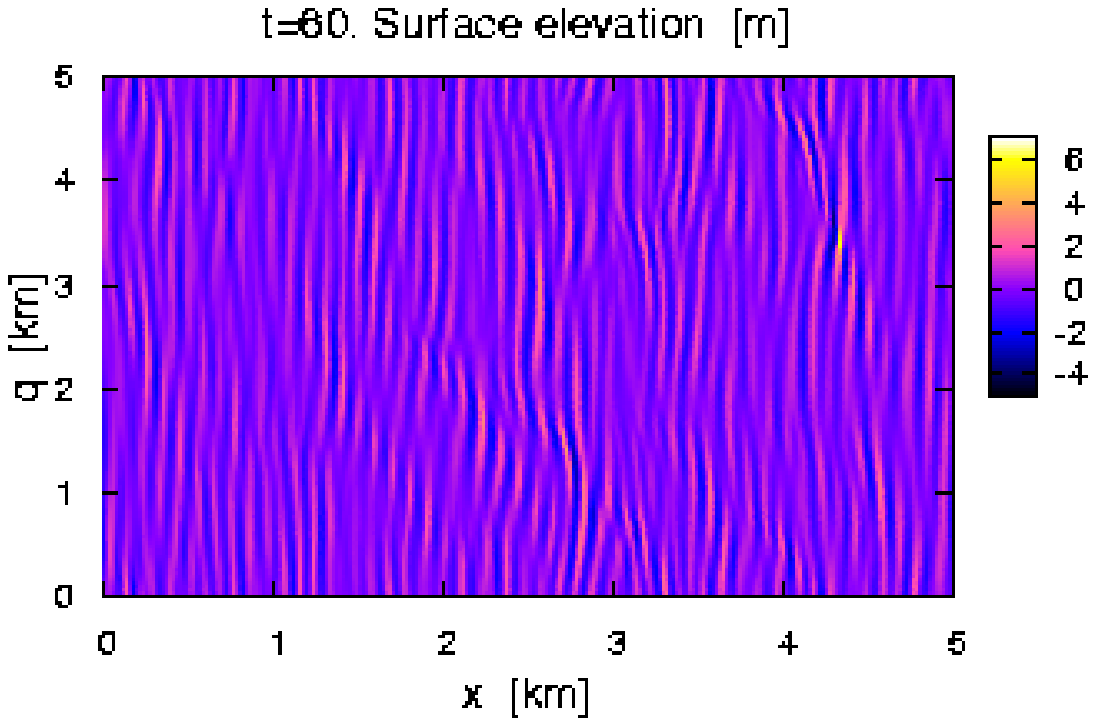,width=84mm}    
\end{center}
\begin{center}
  \epsfig{file=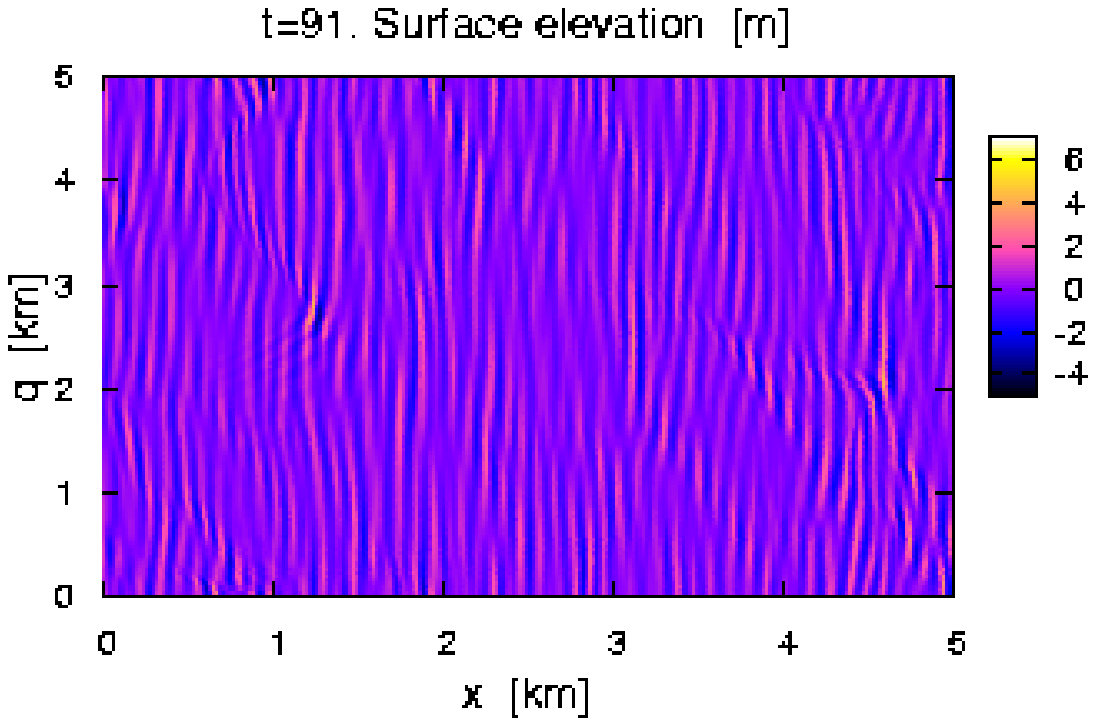,width=84mm}    
\end{center}
\caption{(Color online). 
Top: map of the free surface at $t=50$ (7 min 30 sec). 
The rogue wave has coordinates $x\approx 3.7$ km, $q\approx 3.7$ km.
Middle: map at $t=60$ (9 min 1 sec). The rogue wave is at
$x\approx 4.3$ km, $q\approx 3.4$ km.
Bottom: map at $t=91$ (13 min 40 sec). The rogue wave is at
$x\approx 1.3$ km, $q\approx 2.8$ km, and a specific wave 
pattern behind of it is visible.
} 
\label{t1-t2-t5-maps}
\end{figure}
\begin{figure}
\begin{center} 
  \epsfig{file=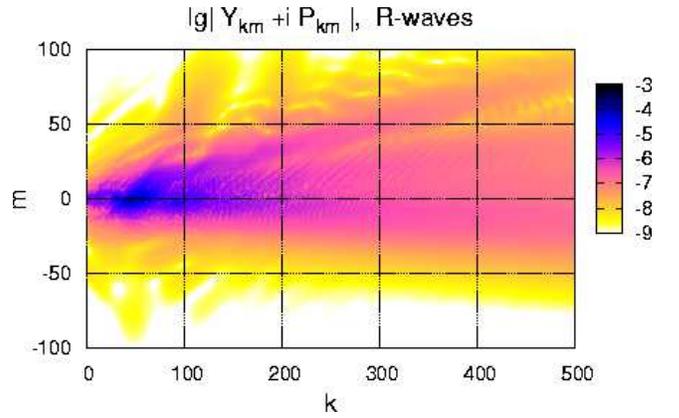,width=85mm}    
\end{center}
\caption{(Color online). Spectrum of the right-propagating waves at $t=50$.
Here shown is $\log_{10}|Y_{km}+iP_{km}|$, where 
$P_{km}=(k^2+m^2)^{1/4}\psi_{km}$.} 
\label{R_micro-50}
\end{figure}
\begin{figure}
\begin{center}
  \epsfig{file=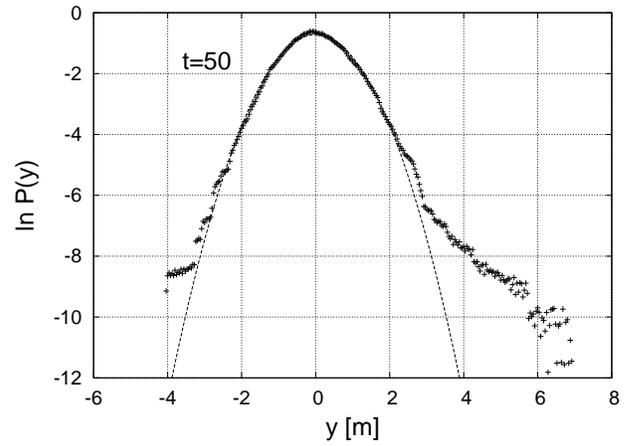,width=85mm}  
\end{center} 
\caption{Distribution of the surface elevation $y(x,q)$ at $t=50$ 
(no averaging over the time is done).} 
\label{Y_distribution}
\end{figure}

Here it is necessary to present the equations that were simulated. 
Their detailed derivation and discussion
can be found in Refs.~\cite{R2005PRE, RD2005PRE}. 
We use Cartesian coordinates $x,q,y$, with $y$ axis up-directed.
The symbol $z$ denotes the complex combination: $z\equiv x+iy$. For every value
of $q$, at any time moment $t$, there exists an analytical function 
$z(u+iv,q,t)$ which determines a conformal mapping of the lower half-plane 
of a complex variable $w=u+iv$ on a region below the free surface. A shape of
the free surface is given in a parametric form:
\begin{equation}\label{Z_Y}
Z= X+iY=z(u,q,t)=u+(i-\hat H)Y(u,q,t).%
%\quad Z_u=1+(i-\hat H)Y_u.
\end{equation} 
The Hilbert operator $\hat H$ is diagonal in the Fourier representation:
it multiplies the Fourier-harmonics 
$$
Y_{km}(t)\equiv\int Y(u,q,t)e^{-iku-imq}du\,dq
$$ 
by $[i\,\mbox{sign\,}k]$,  so that
\begin{equation}\label{HY}
\hat H Y(u,q,t)=\int [i\,\mbox{sign\,}k] Y_{km}(t)e^{iku+imq}
{dk\, dm}/({2\pi})^2.
\end{equation}
Thus, the first unknown function is $Y(u,q,t)$. The second unknown function is
the boundary value $\psi(u,q,t)$ of the velocity potential,
$$
\psi(u,q,t)=\int \psi_{km}(t)e^{iku+imq} {dk\, dm}/({2\pi})^2.
$$
Correspondingly, we have two main equations of motion. 
They are written below in a Hamiltonian
non-canonical form involving the variational derivatives 
$(\delta{\cal K}/\delta\psi)$ and $({\delta{\cal K}}/{\delta Z})$, where
${\cal K}\{\psi,Z,\bar Z\}$ is the kinetic energy.
The first equation is the so called kinematic condition on the free surface:
\begin{equation}\label{kinematic}
Z_t=iZ_u(1+i\hat H )\left[\frac{(\delta{\cal K}/\delta\psi)}
{|Z_u|^2}\right].
\end{equation}
The second equation is the dynamic condition (Bernoulli equation):
\begin{eqnarray}
\psi_t&=&-g\,\mbox{Im\,}Z
-\psi_u\hat H\left[\frac{(\delta{\cal K}/\delta\psi)}{|Z_u|^2}\right]
\nonumber\\
\label{Bernoulli}
&+&\frac{\mbox{Im}\left((1-i\hat H)
\left[2({\delta{\cal K}}/{\delta Z})Z_u
+ ({\delta{\cal K}}/{\delta\psi})\psi_u\right]\right)}{|Z_u|^2},
\end{eqnarray}
where $g$ is the gravitational acceleration.
Equations (\ref{kinematic}) and (\ref{Bernoulli}) completely determine
evolution of the system, provided the kinetic energy functional 
${\cal K}\{\psi,Z,\bar Z\}$ is explicitly given. Unfortunately, in 3D there is
no exact compact expression for ${\cal K}\{\psi,Z,\bar Z\}$. However,
for long-crested waves propagating mainly in the $x$ direction 
(the parameter $\epsilon\sim(l_x/l_q)^2\ll 1$),
we have an approximate kinetic-energy functional in the form
\begin{equation}\label{K_approx}
{\cal K}\approx
\tilde{\cal K}=-\frac{1}{2}\int \psi\hat H\psi_u \,du\,dq +\tilde{\cal F},
\end{equation}
where the first term describes purely 2D flows, and weak 3D corrections
are introduced by the functional $\tilde{\cal F}$:
\begin{eqnarray}
\tilde{\cal F}&=&\frac{i}{8}
\int(Z_u\Psi_q-Z_q\Psi_u)\hat G
\overline{(Z_u\Psi_q-Z_q\Psi_u)}\,du\,dq
\nonumber\\
&+&\frac{i}{16}\int\Bigg\{\left[
(Z_u\Psi_q-Z_q\Psi_u)^2/{Z_u}\right]\,\hat E\overline{(Z-u)} 
\nonumber\\
&&\qquad - (Z-u)\,\hat E\overline{\left[(Z_u\Psi_q-Z_q\Psi_u)^2/{Z_u}\right]}
\Bigg\}\,du\,dq.
\label{H_modified}
\end{eqnarray}
Here $\Psi\equiv (1+i\hat H)\psi$, and
the operators $\hat G$ and  $\hat E$ are diagonal in the Fourier
representation:
\begin{equation}\label{G_def}
G(k,m)=\frac{-2i}{\sqrt{k^2+m^2}+|k|},
\end{equation}
\begin{equation}\label{E_def}
E(k,m)=\frac{2|k|}{\sqrt{k^2+m^2}+|k|}.
\end{equation}
A difference between the above expression (\ref{K_approx})
and the unknown true water-wave kinetic energy functional is of order 
$\epsilon^2$, since $G(k,0)=1/(ik)$ for positive $k$, and $E(k,0)=1$
(see Refs.~\cite{R2005PRE, RD2005PRE}).
Besides that, the linear dispersion relation resulting from 
$\tilde{\cal K}$ is correct in the entire Fourier plane 
(it should be noted that in Ref.~\cite{RD2005PRE} 
another approximate expression for $\tilde{\cal K}$ 
was used, also resulting in the first-order accuracy on $\epsilon$ and 
correct linear dispersion relation). 
Thus, we have 
$({\delta{\cal K}}/{\delta \psi})\approx({\delta\tilde{\cal K}}/{\delta \psi})$
and $({\delta{\cal K}}/{\delta Z})\approx({\delta\tilde{\cal F}}/{\delta Z})$
in equations (\ref{kinematic}-\ref{Bernoulli}), with explicit expressions
closing the system:
\begin{equation}\label{tilde_H_psi}
\frac{\delta\tilde{\cal K}}{\delta\psi}=-\hat H\psi_u+
2\,\mbox{Re\,}\left[(1-i\hat H)\frac{\delta\tilde{\cal F}}{\delta\Psi}
\right],
\end{equation}
%%%%%%%%%%%%%%%%%%%%%%%%%%%%%%%%%%%%%%%%%%%%%%%%%%%%%%%%%%%%%%%%%%%%%%%%%
\begin{eqnarray}
\frac{\delta\tilde{\cal F}}{\delta\Psi}&=&\frac{i}{8}
Z_q\hat\partial_u\Big[\hat G
\overline{(Z_u\Psi_q-Z_q\Psi_u)} \nonumber\\
&&\qquad\qquad+(\Psi_q-Z_q\Psi_u/{Z_u})
\hat E\overline{(Z-u)}\Big]
\nonumber\\
&&-\frac{i}{8}\,Z_u\,\hat\partial_q\Big[
\hat G\overline{(Z_u\Psi_q-Z_q\Psi_u)}\nonumber\\
&&\qquad\qquad + (\Psi_q-Z_q\Psi_u/{Z_u})\hat E\overline{(Z-u)}\Big],
\label{tilde_F_Psi}
\end{eqnarray}
\begin{eqnarray}
\frac{\delta\tilde{\cal F}}{\delta Z}&=&-\frac{i}{8}
\Psi_q\hat\partial_u\Big[\hat G\overline{(Z_u\Psi_q-Z_q\Psi_u)}\nonumber\\
&&\qquad\qquad
+ (\Psi_q-Z_q\Psi_u/{Z_u})\hat E\overline{(Z-u)}\Big]
\nonumber\\
&&+\frac{i}{8}\,
\Psi_u\,\hat\partial_q\Big[
\hat G \overline{(Z_u\Psi_q-Z_q\Psi_u)}\nonumber\\
&&\qquad\qquad+ (\Psi_q-Z_q\Psi_u/{Z_u})\hat E \overline{(Z-u)}\Big]\nonumber\\
&&+\frac{i}{16}\,\Big[\hat\partial_u[(\Psi_q-Z_q\Psi_u/Z_u)^2
\hat E \overline{(Z-u)}]\nonumber\\
&&\qquad\qquad\qquad-\hat E\overline{(\Psi_q-Z_q\Psi_u/{Z_u})^2{Z_u}}
\Big].
\label{tilde_F_Z}
\end{eqnarray}
%%%%%%%%%%%%%%%%%%%%%%%%%%%%%%%%%%%%%%%%%%%%%%%%%%%%%%%%%%%%%%%%%%%%%%%%%

\section{Results of the numerical experiment}
%{\it Results of the numerical experiment.---}
\begin{figure}
\begin{center} 
(a)\epsfig{file=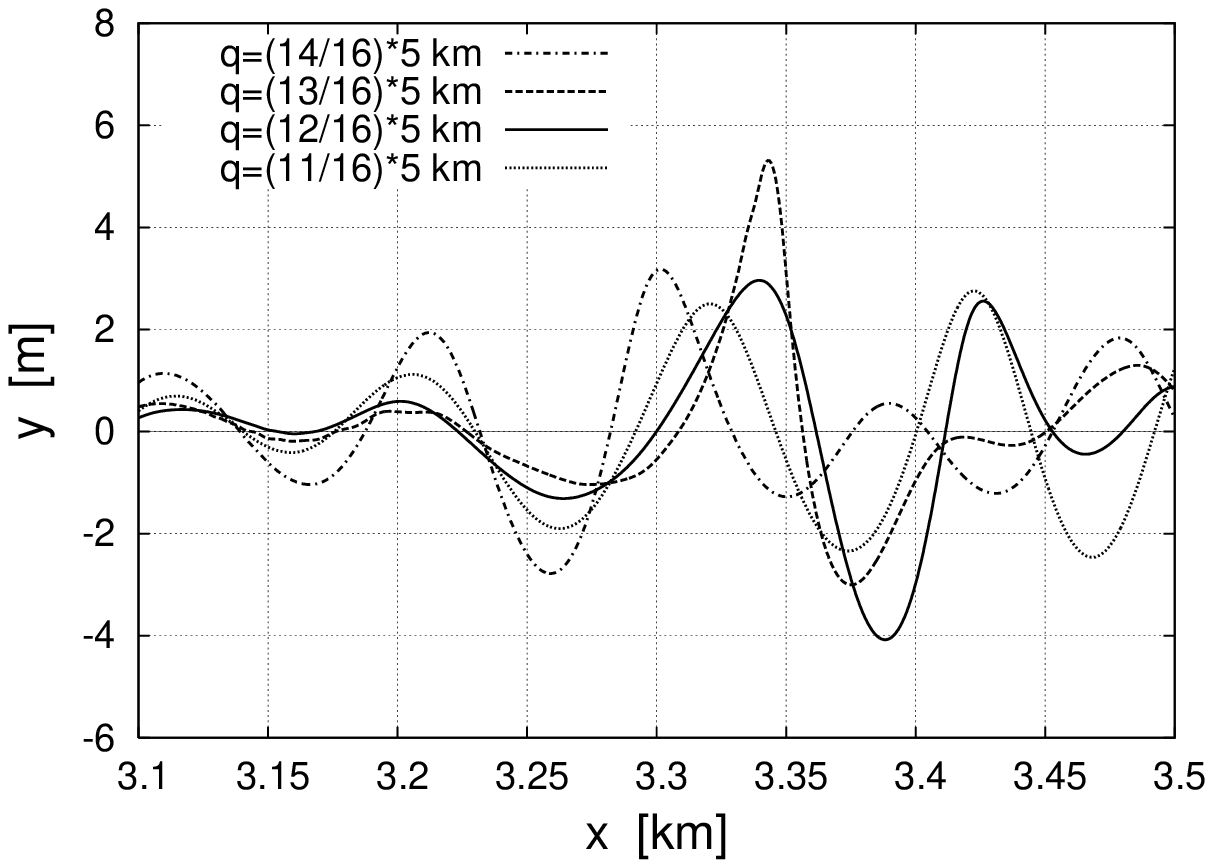,width=70mm}%\epsfig{file=XY-bw-45.eps,width=70mm}  
\end{center} 
\begin{center} 
(b)\epsfig{file=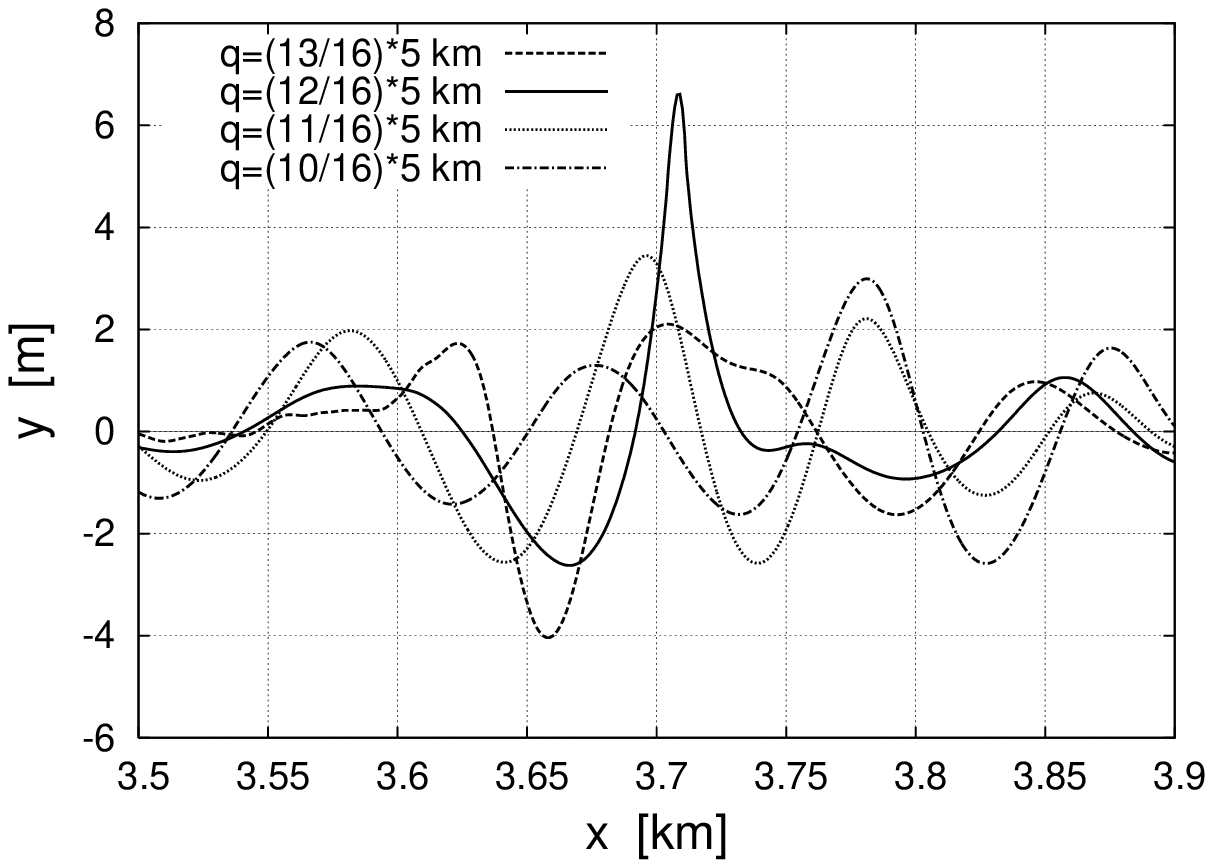,width=70mm}%\epsfig{file=XY-bw-50.eps,width=70mm} 
\end{center} 
\begin{center} 
(c)\epsfig{file=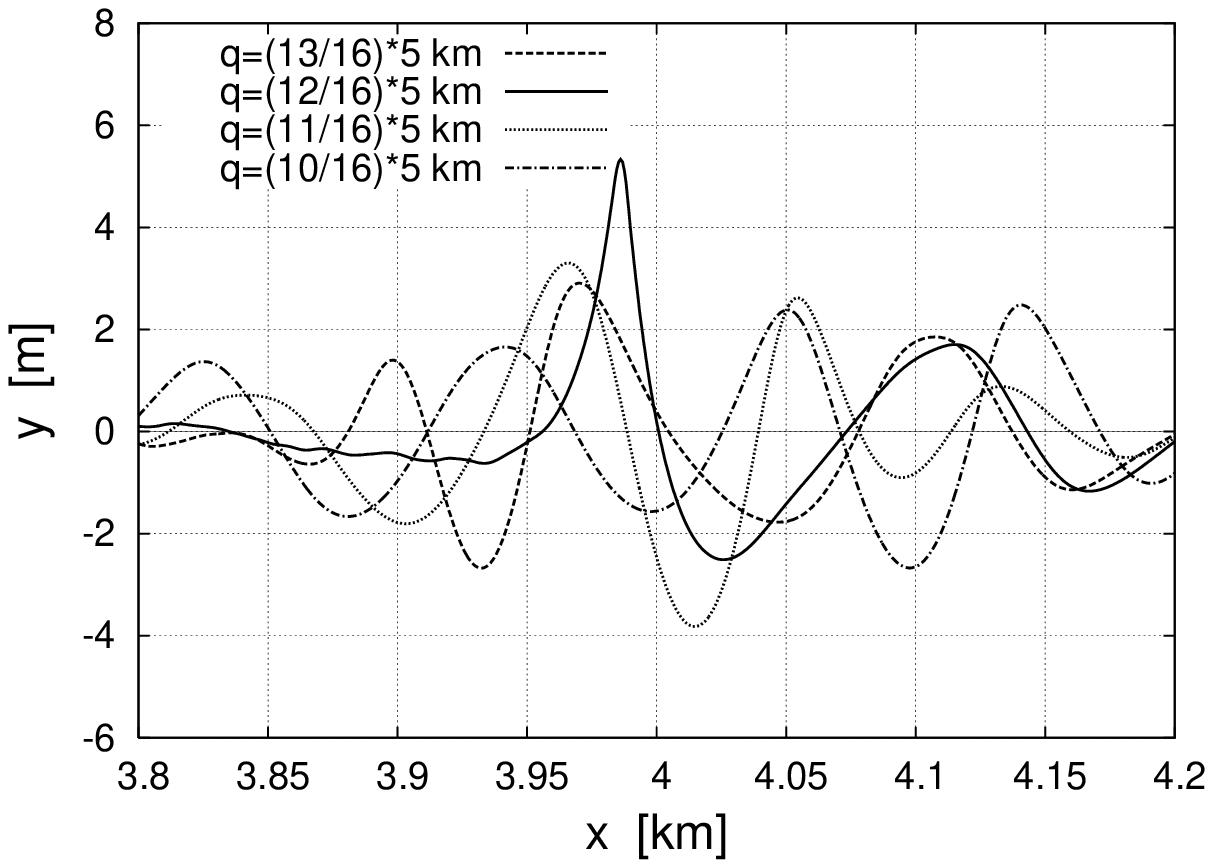,width=70mm}%\epsfig{file=XY-bw-55.eps,width=70mm} 
\end{center} 
\begin{center} 
(d)\epsfig{file=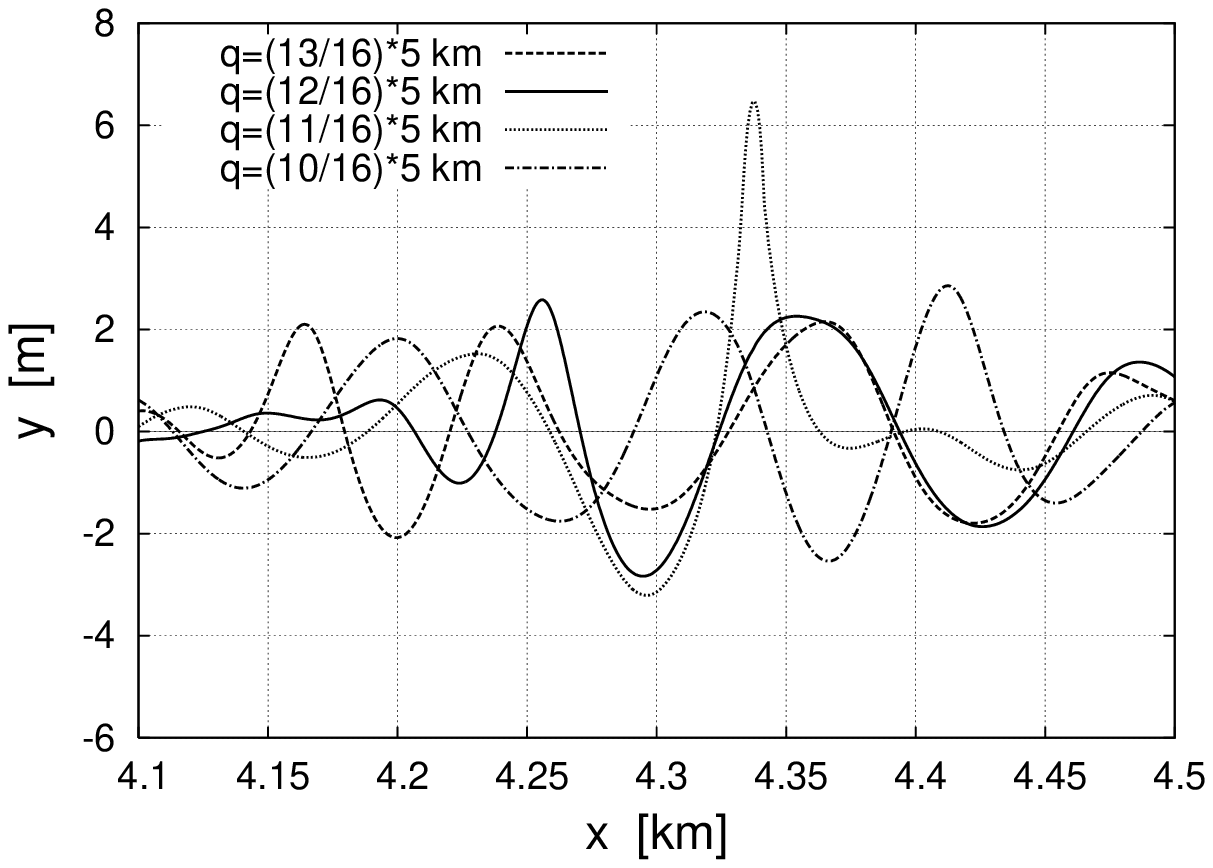,width=70mm}%\epsfig{file=XY-bw-60.eps,width=70mm} 
\end{center}
\caption{Rogue wave profiles at $t=45$, $t=50$, $t=55$, 
and $t=60$.} 
\label{t1-t2-t3-t4-profiles}
\end{figure}

Following the procedure described in Ref.~\cite{RD2005PRE},
a numerical experiment has been performed, which is described below. 
A square $5\times 5$ km in $(u,q)$-plane with periodic boundary conditions
was reduced to the standard square $2\pi\times 2\pi$ and discretized 
by $N\times L$ points. Thus, all the wave numbers $k$ and $m $ are integer. 
Dimensionless time units imply $g=1$.
As an initial state, a superposition of quasi-randomly placed wave packets 
was taken, with 25 packets having wave vector $(60,2)$, 25 packets having
wave vector $(50,0)$, 16 packets with $(40,-2)$, and 12 packets with 
$(30,1)$. Amplitudes of the packets with $k=50$ were dominating.
Thus, a typical wave length was 100 m, and a typical dimensionless wave period 
$T=2\pi/\sqrt{50}\approx 1$.
The crest of the highest wave was initially less than 3 m above zero level.
A map of the free surface at $t=0$ is shown in Fig.\ref{t0-map}.
It is clear from this figure that initially $\epsilon\sim 0.01$. 

The evolution of the system was computed with $N=16384$ and $L=256$ 
to $t=40.0$, until beginning of a rogue wave formation. After  $t=40.0$, 
the rogue wave was present in the system (see Fig.~\ref{t1-t2-t5-maps}),
and during many wave periods its height in maximum 
was approximately $7$ m, as Fig.~\ref{Ymax} shows. 
It resulted in widening of the wave spectrum (see Fig.~\ref{R_micro-50}, where
$\epsilon\sim m^2/k^2\sim 0.05$), 
and therefore $L=512$ was employed from $t=40.0$ to $t=60.0$. 
Within this period, the total energy was decreased by $0.5\%$ 
due to numerical errors.
Finally, from $t=60.0$ and to the end of the experiment, $L=1024$ was used
to avoid progressive loss of the accuracy (the last stage has required computer 
with 3072M memory, and it took 5 min per one step of integration).

%\clearpage

The presence of rogue wave strongly affects the probability distribution
function $P(y)$ of the free surface elevation $y(x,q)$. 
Fig.~\ref{Y_distribution}  shows that the distribution has a Gaussian core 
and ``heavy'' tails, which are not symmetric -- large positive $y$ are more
probable than large negative $y$.

The most interesting observation of the present numerical experiment is that
the freak wave can exist for a relatively long time without significant 
tendency towards breaking or disappearing. While ``living'', 
the big wave does something similar to breathing, 
as shown in Fig.~\ref{t1-t2-t3-t4-profiles}. The rogue wave
propagates along the free surface (with the typical group velocity, 
but there is also a displacement in $q$-direction), and position of the
highest crest is alternately ahead or behind of the deepest trough.
Very roughly this behavior corresponds to a short wave envelope 
(with approximately one wave length inside) filled with a strongly nonlinear
periodic Stokes-like wave.  The time period of this ``breathing'' roughly 
equals to two wave periods, which property seems natural due to the fact that 
the group velocity of the gravitational waves is one half of the phase velocity.
After 11 periods of ``breathing'' with the almost constant amplitude 7 m,
the rogue wave gradually irradiates accumulated energy into a specific
wave pattern visible in Fig.~\ref{t1-t2-t5-maps} at $t=91$.
This wave pattern nearly corresponds to the resonance condition
$$
\omega({\bf k}_0) +
{\bf V}_{gr}({\bf k}_0)\cdot({\bf k}-{\bf k}_0)-\omega({\bf k})=0.
$$
where the wave vector ${\bf k}_0=(k_0,m_0)$ characterizes the rogue wave, 
and $\omega({\bf k})=(g|{\bf k}|)^{1/2}$ is the linear dispersion relation.
However, a more accurate explanation and an analytical study of the observed 
coherent nonlinear structure is a subject of future work.

\section{Summary}

Thus, the recently developed fully nonlinear theory for long-crested water
waves, together with the corresponding FFT-based numerical 
method \cite{RD2005PRE} are shown in this work to be an adequate tool 
for modeling rogue waves in close to real situations, that is with many random
waves propagating mainly along a definite horizontal direction. 
Now it has been possible to deal with quite high spatial resolutions, 
since in the present algorithm all the non-local operations are reduced 
to the FFT computing, and the latter is really fast with modern numerical
libraries. Different dynamical regimes of the rogue wave evolution can be
investigated. In particular, the present article reports observation of a
long-lived rogue wave. Such waves are definitely important from practical
viewpoint.

\end{document}